\documentclass[aps,nofootinbib,twocolumn,showpacs,preprintnumbers]{revtex4}

\usepackage{citesort,latexsym,ifthen,graphics,color}

\newcommand{\ie}{{i.e.}}
\newcommand{\cf}{{cf.}}
\newcommand{\eg}{{e.g.}}
\newcommand{\aka}{{a.k.a.}}
\newcommand{\etc}{{etc.}}

\newcommand{\lhs}{left-hand side}
\newcommand{\rhs}{right-hand side}

\newcommand{\be}{\begin{equation}}
\newcommand{\ee}{\end{equation}}
\newcommand{\bea}{\begin{eqnarray}}
\newcommand{\eea}{\end{eqnarray}}
\newcommand{\beas}{\begin{eqnarray*}}
\newcommand{\eeas}{\end{eqnarray*}}

\newcommand{\bear}{\begin{array}{l}}
\newcommand{\eear}{\end{array}}

\newcommand{\bcf}{\begin{center}\begin{figure}}
\newcommand{\ecf}{\end{figure}\end{center}}

\newcommand{\bct}{\begin{center}\begin{table}}
\newcommand{\ect}{\end{table}\end{center}}

\def\Q{{\cal Q}}

\def\eq#1{(\ref{eq:#1})}

\def\Eqn#1{Equation~(\ref{eq:#1})}

\def\eqs#1#2{(\ref{eq:#1}) and~(\ref{eq:#2})}

\def\fig#1{figure~\ref{fig:#1}}

\newcommand{\der}[2]{\ensuremath{\frac{d #1}{d #2}}}

\newcommand{\fder}[2]{\ensuremath{\frac{\delta #1}{\delta #2}}}

\newcommand{\hf}{\frac{1}{2}}

\newcommand{\expectation}[1]{\left\langle #1 \right\rangle}
\newcommand{\measure}[1]{\mathcal{D} #1 \, }

\newcommand{\SU}{\mathrm{SU}}

\def\dd{\dot{\Delta}}

\def\hS{\hat{S}}
\def\e#1{\,{\rm e}^{\displaystyle #1}}
\def\one{\hbox{1\kern-.8mm l}}

\newcommand{\itp}{\Delta^{\! -1}}
\newcommand{\ctp}{(\itp)}
\newcommand{\Ker}[1]{\{#1\}}

\newcommand{\GR}{\cdeps{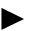}}
\newcommand{\GRk}{\rhd}
\newcommand{\GRkpr}{>}

\newcommand{\DummyKernel}{\ensuremath{\stackrel{\bullet}{\mbox{\rule{1cm}{.2mm}}}}}

\newcommand{\bigdot}[1]{\stackrel{\bullet}{#1}}
\newcommand{\flow}{\Lambda \partial_\Lambda}

\newcommand{\dec}[3][0]{\ensuremath{\left[ #2 \hspace{#1em} \right]^{#3}}}
\newcommand{\norm}{\ensuremath{\Upsilon}}
\newcommand{\GIO}{\mathcal{O}}

\newcommand{\cdeps}[1]{\ensuremath{\begin{array}{c}\includegraphics{#1.eps} \end{array}}} 

\newlength{\VertexWidth}

\newcommand{\OVertex}[1]{
\ensuremath{
	\begin{array}{c}
	\settowidth{\VertexWidth}{$#1$}
	\setlength{\unitlength}{1.2\VertexWidth}
	\begin{picture}(1,1)(0,0)
		\put(0,0){\framebox(1,1){$#1$}}
	\end{picture}
	\end{array}
}
}

\newcommand{\OTower}{
	\begin{array}{c}
	\vspace{1ex}
		\OVertex{\, \GIO \,}
	\\
		\dec{
			\begin{array}{c}\includegraphics{ReducedWEA.epsi}\end{array}
		}{j}
	\end{array}
}

\newcommand{\jhep}[3]{{JHEP} #1 (#2) #3}
\newcommand{\NuclPhys}[4]{{Nucl.\ Phys.\ }\textbf{#1 #2} (#3) #4}
\newcommand{\PhysRev}[4]{{Phys.\ Rev.\ }\textbf{#1 #2} (#3) #4}
\newcommand{\IntJModPhys}[4]{{Int.\ J.\ Mod.\ Phys.\ }\textbf{#1 #2} (#3) #4}
\newcommand{\PhysRep}[4]{{Phys.\ Rep.\ }\textbf{#1 #2} (#3) #4}
\newcommand{\PhysRept}[3]{{Phys.\ Rept.\ }\textbf{#1} (#2) #3}

\newcommand{\ProgTheorPhys}[3]{{Prog.\ Theor.\ Phys.\ }\textbf{#1} (#2) #3}
\newcommand{\ProgTheorPhysS}[3]{{Prog.\ Theor.\ Phys.\ Suppl.\ }\textbf{#1} (#2) #3}

\newcommand{\CEurJPhys}[3]{{Central Eur.\ J.\ Phys.\ }\textbf{#1} (#2) #3}

\newcommand{\arxiv}[1]{#1}
\newcommand{\hepth}[1]{hep-th/#1}
\newcommand{\hepph}[1]{hep-ph/#1}

\newcommand{\RevModPhys}[3]{{Rev.\ Mod.\ Phys.\ }\textbf{#1} (#2) #3}

\newcommand{\jphysa}[3]{J.\ Phys.\ {\bf A} #1 (#2) #3}

\begin{document}

\title{Universality From Very General Nonperturbative Flow Equations in QCD}

\author{
	Oliver J.~Rosten
}

\affiliation{Dublin Institute for Advanced Studies, 10 Burlington Road, Dublin 4, Ireland}
\email{orosten@stp.dias.ie}

\begin{abstract}
	In the context of very general exact 
	renormalization groups,
	it will be shown that, given a vertex
	expansion of the Wilsonian effective action,
	remarkable progress can be made without making 
	any approximations. Working in QCD 
	we will derive, in a manifestly gauge invariant
	way, an exact diagrammatic expression
	for the expectation value of an
	arbitrary gauge invariant operator,
	in which many of the non-universal
	details of the setup do not explicitly
	appear. This provides a new
	starting point for attacking 
	nonperturbative problems.
\end{abstract}

\preprint{DIAS-STP-06-22}
\maketitle

In this paper, we will describe a major development
in the understanding of how a manifestly gauge invariant approach
to QCD, based on the exact renormalization group (ERG)~\cite{ERG},
may be used as a practical 
tool for extracting nonperturbative information.

Given some field theory,
the basic idea of the ERG 
(for reviews see~\cite{Reviews,TRM-elements,Pawlowski:2005xe})
is the implementation
of a momentum cutoff, $\Lambda$,
in such a way that the physics
at this scale
is described in terms of parameters relevant
to this scale. The effects of the modes above
$\Lambda$, which have been integrated out,
are encoded in the Wilsonian effective action,
$S_\Lambda$.
The ERG (or flow) equation determines how $S_\Lambda$ evolves
with $\Lambda$, thereby linking physics at different
energy scales and so providing
access to nonperturbative physics. One of the great benefits
of the ERG approach is that renormalization is built in:
solutions to the flow equation,
from which physics can be extracted,
are naturally phrased directly in terms of renormalized parameters.

Compared to alternative ERG approaches to QCD
(for a comprehensive review of these, 
see~\cite{Pawlowski:2005xe}), the 
manifestly gauge invariant approach advocated
by this paper has both a major advantage and a
major disadvantage. The benefit of the formalism is
that gauge invariance is manifestly preserved along
the entire flow: not only are the 
Ward-Takahashi Identities (WTIs) not modified (as
is generically the case), but they are in fact
particularly simple as a consequence of the
exact preservation of gauge invariance. The
drawback of the approach is that the implementation
of a manifestly gauge invariant 
ERG~\cite{ymi,aprop,mgierg1,QCD} requires considerable
complication.
A pressing question
for this manifestly gauge invariant approach, then,
is whether the complications can be reduced to a level
where the elegance and power of manifest gauge invariance
provides a real, practical advantage.

Already, there have been tantalizing hints that this might
be possible, as evidenced by the surprisingly high
degree of universality found at intermediate stages
of perturbative computations of $\beta$ function 
coefficients~\cite{mgierg1,mgierg2,RG2005,mgiuc,QCD} and
expectation values of gauge invariant operators~\cite{evalues}.
By this, we mean the following.
The complications arising in the implementation of
a manifestly gauge invariant ERG 
reside in the regularization and suitable generality
of the way in 
which the high energy modes are integrated out;
as such, the complications amount
to nonuniversal details of which
physical observables must be independent.
Now, despite both the
$\beta$ function
and the expectation values of generic gauge invariant
operators being scheme dependent\footnote{Of course,
in certain schemes the first two perturbative
$\beta$-function coefficients are universal.}
it is found that many (though of course not all) of
the explicit nonuniversalities nevertheless
cancel out, 
to all orders in perturbation theory.
In this paper, we will show how these cancellations 
can be extended nonperturbatively, which
represents a
crucial first step in understanding how to 
\emph{practically}
use this formalism
for nonperturbative calculations. 

We now illustrate
these considerations in the context of expectation
values of gauge invariant operators, whose flow can
be computed by taking the ERG equation and
considering linear, gauge invariant 
perturbations to the Wilsonian effective 
action~\cite{evalues}.
Thus, having inserted a gauge invariant 
operator, $\GIO$, at the bare scale, we
can track its evolution as we integrate out degrees
of freedom.
In the limit that all fluctuations have been integrated
out, the value of this (effective) operator, 
$\GIO_{\Lambda=0}$, simply
corresponds to the renormalized expectation value we are seeking to
compute:
\be
\label{eq:GIO-zero}
	\expectation{\GIO}_R = \GIO_{\Lambda =0}.
\ee

In~\cite{evalues}, a manifestly gauge invariant,
perturbative expression for the $n$-loop contribution
to the field independent part of
$\GIO_{\Lambda}$---which is the only contribution that
survives the limit $\Lambda \rightarrow 0$---, 
$g^{2(n-1)}\GIO_n$, 
was derived, such that%
\footnote{
The coupling, $g$, is scaled out of the covariant derivative, as a
result of which $S \rightarrow S/g^2$ (similarly $\GIO$).}
\be
\label{eq:EV-Template}
	\der{}{\Lambda} \sum_n \left[g^{2(n-1)} \GIO_n \right] = 
	\der{}{\Lambda} \sum_n \left[g^{2(n-1)}\bar{\Q}_n \right],
\ee
where $\bar{\Q}_n$ is a set of
$n$-loop ERG diagrams.
The expression~\eq{EV-Template}
can be directly integrated
and so we see that the perturbative
contributions to $\expectation{\GIO}_R$ are
given by sets of terms evaluated at the
bare scale, and sets of terms evaluated
at $\Lambda=0$. In the perturbative treatment,
the latter terms in fact vanish~\cite{evalues}
and so $\expectation{\GIO}_R$ can be expressed solely
as contributions directly fixed by the boundary
condition at the bare scale
(at the very end of a calculation, the bare scale
is tuned to infinity, which essentially corresponds to taking
the continuum limit~\cite{TRM-elements}).

Now, the set of diagrams contributing to~\eq{EV-Template}
exhibit a strong degree of universality, in the sense that
many of the nonuniversal details of the setup do not explicitly
appear (there is, of course, still some scheme dependence
present, as should be expected). To understand this statement
better, we now review the nonuniversalities inherent in our
ERG approach and discuss how they generically 
cancel out. 

One of the key ingredients of any
flow equation is that the partition function
(and hence the physics derived from it)
is invariant under the flow. As a consequence of this, the family of
flow equations for some generic fields, $\varphi$, follows
from~\cite{jose}
\be
\label{eq:blocked}
-\flow \e{-S[\varphi]} =  \int_x \fder{}{\varphi(x)} \left(\Psi_x[\varphi] \e{-S[\varphi]}\right),
\ee
where the $\Lambda$ derivative
is performed at constant $\varphi$,
any Lorentz indices \etc\ have been suppressed
and we have written $S_\Lambda$ as just $S$.
The total derivative on
the \rhs\ ensures that the 
partition function $Z = \int \measure{\varphi} \! \e{-S}$
is invariant under the flow.

The primary source of nonuniversal details 
is $\Psi$, which 
parametrizes (the continuum version of a)
general Kadanoff blocking~\cite{Kadanoff} in the 
continuum, and for which we take the
following form~\cite{mgierg1}:
\be
\label{eq:Psi}
	\Psi_x = \hf \Ker{\dd^{\varphi \varphi}(x,y)} \fder{\Sigma}{\varphi(y)},
\ee
where it is understood that we sum over all the elements
of the set of fields $\varphi$.
We now describe each of the components on the \rhs\ of~\eq{Psi}. 
First, there are the ERG kernels,
$\dd^{\varphi \varphi}$, which are generally
different for each of the elements of $\varphi$. Each kernel
incorporates a cutoff function which provides ultraviolet (UV)
regularization. The notation $\Ker{\dd}$ denotes a covariantization
of the kernel which may be necessary, depending on the
symmetries of the theory. Indeed, it is
apparent from~\eqs{blocked}{Psi} that the kernel
essentially ties together two functional derivatives
at points $x$ and
$y$; in gauge theory, we can covariantize this statement
by using \eg\ straight Wilson lines between these
two points. In practice, we leave the covariantization
unspecified, demanding only that it satisfies general
requirements~\cite{ymi,mgierg1}. The remaining ingredient
in~\eq{Psi} is $\Sigma \equiv S - 2\hS$,
where $\hS$
is the seed action~\cite{mgierg1,mgierg2,seed}:
a functional with the same
symmetries as the Wilsonian effective 
action which partially parametrizes how modes are
integrated out along the flow.

The other source of nonuniversality in~\eq{blocked}
is more subtle: it might be necessary to include
some unphysical regulator fields in the set $\varphi$.
Indeed, this is precisely the case in the 
manifestly gauge invariant ERG
formulation of QCD that we employ, where
covariantization of the cutoff functions is
not sufficient to completely regularize the
theory. To furnish a complete 
regularization of QCD
we embed the physical theory into a spontaneously
broken $\SU(N|N)$ gauge theory, with the unphysical
fields supplying precisely the necessary extra 
regularization~\cite{sunn,QCD}. 

Substituting~\eq{Psi} into~\eq{blocked} we perform
the $\Lambda$-derivative on the \lhs. Identifying
terms with the same number of fields on both sides
allows us to write down a flow equation for the
\emph{vertex coefficient functions} of the Wilsonian
effective action (\ie\ the fields themselves and
any symmetry factors have been stripped off). 
Anticipating specialization to QCD, we recall
that, in this case, it is convenient to scale
the coupling out of the covariant derivative~\cite{ymi}.
This has the effect of causing $S \rightarrow S/g^2$
and so we make the replacement $\Sigma \rightarrow \Sigma_g \equiv g^2 S- 2\hS$.
The flow equation for QCD is shown in \fig{Flow}~\cite{QCD}.
\bcf[h]
\resizebox{8.5cm}{!}{\includegraphics{figure1.epsi}}
\caption{The diagrammatic form of the QCD
flow equation for vertices
of the Wilsonian effective action.}
\label{fig:Flow}
\ecf

The first term on the \lhs\ represents
the flow of all independent
Wilsonian effective action
vertex coefficient functions corresponding
to the set of fields, $\{f\}$.
Our notation is slightly different
to previous works for later clarity: 
since the $\Lambda$-derivative
strikes just a vertex coefficient 
function---all fields
having been stripped off---we need not
write this as a partial derivative with
fields held constant (\cf~\eq{blocked}).
The term $\sum_{\chi \epsilon \{f\}} \gamma^{(\chi)}$
explicitly takes account of the anomalous
dimensions of the fields which suffer
field strength renormalization.
The field $\chi$ belongs to the set of fields 
$\{f\}$ and
the notation $\gamma^{(\chi)}$
just stands for the anomalous dimension of
the field $\chi$ (which is zero for all
but the components of the superfields into
which the physical quark fields are embedded~\cite{QCD}).

The lobes on the \rhs\ of the flow equation
are vertex coefficient functions of
$S$ and $\Sigma_g$. These lobes are joined together
by vertices of the covariantized ERG kernels,
denoted by \DummyKernel, which, like the action vertices, can be 
decorated by fields belonging to $\{f\}$.
The rule for decorating the diagrams on
the \rhs\ is simple: the set of fields, $\{f\}$, are distributed in 
all independent ways between the component objects of each diagram.
Embedded within the diagrammatic rules is a prescription for evaluating the
group theory factors, for which the reader is referred
to~\cite{mgierg1}.

Now, returning to~\eq{EV-Template}, it is the case
that the set of diagrams on the \rhs\ contains no explicit
dependence on either the seed action or the details
of the covariantization of the kernels. Before stating
what the \rhs\ does depend on, we describe
the procedure by which the aforementioned nonuniversal
details cancel out. The key is to utilize the 
(perturbative) diagrammatic
calculus, proposed in~\cite{aprop}, refined in~\cite{mgierg1,RG2005}
and completed in~\cite{mgiuc}.

The central ingredient to this calculus is the `effective
propagator relation'~\cite{aprop,mgierg1,seed}, which arises as follows.
First, we perform a perturbative expansion of the actions
and flow equation. Secondly, 
we turn the freedom inherent in the seed action to our advantage
by setting the seed action classical, two-point vertices equal to their
Wilsonian effective action counterparts.
It then follows, from
the classical part of the flow equation, that sets of these vertices are related
to sets of kernels according to:
\be
\label{eq:Prototype-EPR}
	S_{0 M R}^{\ \; \varphi \; \varphi}(p) \Delta^{\,\varphi\,\varphi}_{RN}(p^2/\Lambda^2) = \delta_{MN} - p'_M p_N,
\ee
where it is understood that we sum not only over the index, $R$, but also
over the elements of $\varphi$ which carry this index.
$S_{0 M R}^{\ \;\varphi\;\varphi}(p)$ is a classical, 
two-point vertex carrying
indices $M$ and $R$ and momenta $p_\mu$ and $-p_\rho$. 
$\Delta^{\,\varphi\,\varphi}_{RN}(p^2/\Lambda^2)$ is an integrated kernel%
\footnote{Generically, the (integrated) kernels in~\eq{Prototype-EPR}
are (integrated) linear combinations of the kernels which appear in the
flow equation.}
\aka\ effective propagator
\[
	\bigdot{\Delta} \equiv -\flow \Delta,
\]
where the $\Lambda$-derivative is performed with any
dimensionless couplings on which $\Delta$ depends
held constant~\cite{mgierg1,mgierg2,QCD}.
On the \rhs\ of~\eq{Prototype-EPR}
there is a Kronecker-$\delta$ function and
a remainder term comprising a function of $p_\mu$, $p'_M$,
and a function of $p_\nu$, $p_N$. In QCD, these functions
are non-null in the gauge sector (which we recall is extended
due to the embedding into $\SU(N|N)$). For this reason,
the remainders are referred to as `gauge remainders'. As
an example, in the physical gauge
sector---which has gauge field $A^1_\mu$---, the 
relationship~\eq{Prototype-EPR} is
\[
	S^{\ A^1 A^1}_{0 \mu \ \; \alpha} (p) \Delta^{A^1 A^1}_{\alpha \ \, \nu} (p) = \delta_{\mu\nu} - \frac{p_\mu}{p^2} p_\nu.
\]
From this we see that the effective propagator is the inverse of
the classical, two-point vertex only in the transverse space.
It is important to note that 
the effective propagators are by no means propagators
in the usual sense; indeed, they cannot be, since we
have not fixed the gauge. However, they have a similar
form and play a similar diagrammatic role, and so the terminology
`effective propagator' is intuitively helpful.
The components of the gauge remainder are identified as
follows: $p'_M = p_\mu / p^2$, $p_N = p_\nu$.

\Eqn{Prototype-EPR} has a very useful diagrammatic representation,
shown below:
\[
\resizebox{8.5cm}{!}{\includegraphics{EPR.epsi}}
\]

We have attached the effective propagator,
denoted by a solid line, 
to an arbitrary structure since it only
ever appears as an internal line.
The field labeled by $M$ can be any of the physical
or regulator fields. 
The object $\GR \!\! \equiv \; \GRkpr \!\!\! \GRk$ 
is a gauge remainder. Recalling~\eq{Prototype-EPR},
we identify $\GRkpr$ with $p'_M$ and $\GRk$ with $p_N$.

The reason that the effective propagator relation
is so useful is that it allows diagrams to be simplified:
in any term where a classical, two-point vertex
is attached to an effective propagator, we can
collapse this structure down to  Kronecker-$\delta$
and a gauge remainder. When deriving~\eq{EV-Template}
we find that the diagrams formed by
the Kronecker-$\delta$ contribution
cancel against
other terms. This leaves the diagrams containing
gauge remainders, which it turns out can be 
further processed by using the WTIs~\cite{ymi,aprop,mgierg1}.
In a subset of the resulting diagrams
the effective propagator relation can again be
applied. Iterating
the procedure, we find that all explicit dependence
on the seed action and details of the covariantization
of the kernels cancels out. When the dust
has settled, the set of diagrams contributing
to $\expectation{\GIO}_R$ comprise only four ingredients,
the first three of which are vertices of
$\GIO_\Lambda$, effective propagators and
instances of the gauge remainder component
$\GRkpr$. The final ingredient is
vertices of the Wilsonian
effective action, but where none of these 
are classical, two-point vertices. This
is because
all such terms have been removed by
application of the effective propagator relation.
This has led us, in the past~\cite{RG2005,mgiuc},
to introduce reduced vertices, defined as follows: 
given the loop expansion of
a  Wilsonian effective action (or seed action)
vertex with an arbitrary number of legs
we subtract off the classical, two-point component.

The major breakthrough of this paper is the
realization that the diagrammatic calculus
has a non-perturbative extension. The apparent
barrier to this is that
the effective propagator relation and the reduction
of the Wilsonian effective action vertices are
apparently rooted in perturbation theory, since both involve
the introduction and utilization of classical
vertices.
The solution to this problem
is as simple as it is obvious: we \emph{define}
a set of two-point vertices $\ctp^{\,\varphi \, \varphi}_{S \,T}(p)$,
such that
\be
\label{eq:GenEP}
	\ctp^{\; \varphi\;\varphi}_{MR}(p) \Delta^{\,\varphi\,\varphi}_{RN}(p^2/\Lambda^2) = \delta_{MN} - p'_M p_N.
\ee
Clearly, $\ctp^{\; \varphi\;\varphi}_{MR}(p)$ is just numerically
equal to $S_{0 M R}^{\ \; \varphi\;\varphi}(p)$, but it makes
sense to isolate any instances of $\ctp^{\;\varphi \;\varphi}_{MR}(p)$
in  action vertices even when no
loop expansion has been performed. 
Essentially,  all
we have done is change notation to emphasise
that the objects of which the integrated kernels
are the inverses (in the appropriate space)
can be introduced independently of performing
a perturbative expansion. This
subtle shift of perspective holds
the key to extending the diagrammatic 
calculus nonperturbatively.
%
%
The complimentary part of this strategy is to
generalize the reduction of
action vertices according to:
\[
\resizebox{4cm}{!}{\includegraphics{RWEA.epsi}},\
\resizebox{4cm}{!}{\includegraphics{RSA.epsi}},
\]
where it is understood that the vertex with argument $\itp$
is null unless the number of fields in the set $\{f\}$
is precisely equal to two.
In the weak coupling limit, 
these definitions just reduce to those of~\cite{RG2005,mgiuc}:
namely that a reduced vertex does not possess a
classical, two-point component.

Remarkably, the introduction of the set of vertices,
$\ctp^{\, \varphi\,\varphi}_{S\,T}(p)$, together with the 
generalization
of the reduced vertices are the only steps necessary to
apply the diagrammatic calculus nonperturbatively. Thus,
it turns out
that we can write, \emph{exactly},
\be
\label{eq:EV-Template2}
	\der{\GIO}{\Lambda} = \der{\bar{\Q}}{\Lambda},
\ee
where we take $\GIO$ to represent the field independent
part of $\GIO_\Lambda$. 
The set of diagrams, $\bar{\Q}$, is given by
\be
\label{eq:Qbar}
\bar{\Q} \equiv 
- 2\sum_{s=1}^\infty \sum_{m=0}^{2s-1} \sum_{j=0}^s 
\frac{\norm_{s,j} }{m!} g^{2s}
\dec{
  \OTower
}{\Delta^s \GRkpr^m},
\ee
with, for non-negative integers $a$ and $b$,
the definition
\[
	\norm_{a,b} \equiv \frac{(-1)^{b+1}}{a!b!} \left(\frac{1}{2}\right)^{a+1}.
\]

We understand the notation of~\eq{Qbar} 
as follows. The diagrammatic function 
$\bar{\Q}$ stands for all connected diagrams
which can be created from a single vertex of $\GIO_\Lambda$
(with any number of legs),
$j$ reduced Wilsonian effective action vertices
(each with any number of legs),
$s$ effective propagators and $m$ of the gauge remainder
components, $\GRkpr$. For the rules specifying how
to explicitly construct fully fleshed out
diagrams contributing to $\bar{\Q}$ see~\cite{mgiuc,RG2005}.
Notice that the overall $g$ dependence of $\bar{\Q}$
comes
not just from the factor of $g^{2s}$ sitting in front
of the sum over diagrams but also from the vertices
of both $\GIO_\Lambda$ and the Wilsonian effective action;
this is crucial as it is the nonperturbative contributions
to these vertices which will provide the nonperturbative
contributions to $\bar{\Q}$.

Again, we integrate~\eq{EV-Template2} between $\Lambda=0$
and the bare scale. The latter contributions,
for which the coupling is small, contain the
perturbative contributions, \eq{EV-Template},
and additional, nonperturbative parts, arising from
the fact that the vertices appearing in $\bar{\Q}$ 
are exact. The investigation of these contributions
is left to the future; similarly, we defer answering the
question as to
whether the contributions
from $\Lambda=0$ survive in this nonperturbative 
formulation. However, we do note
that, to answer this, we must find out whether the coupling
grows sufficiently fast in the infrared to prevent
the $\Lambda=0$ contributions from vanishing~\cite{evalues}. 
The investigation of this will be greatly helped
by the fact that
the nonperturbative generalization
of the diagrammatic calculus enables us
to derive an exact diagrammatic expression
for the $\beta$ function, with no explicit
dependence on either the seed action, or
details of the covariantization of the kernels.

\begin{acknowledgments}

It is a pleasure to thank Tim Morris and 
Daniel Litim for helpful discussions.
I acknowledge financial support from IrcSet.

\end{acknowledgments}

\end{document}